\documentclass[twocolumn,amssymb, amsmath, hyperref, aps, prd, nofootinbib]{revtex4}
\usepackage{bm}
\usepackage{graphicx}
\expandafter\ifx\csname package@font\endcsname\relax\else
 \expandafter\expandafter
 \expandafter\usepackage
 \expandafter\expandafter
 \expandafter{\csname package@font\endcsname}
\fi
\def\sub#1{_{\mbox{\scriptsize{#1}}}}
\def\ie{\textit{i.e.} }
\def\etal{\textit{et al.} }
\def\eg{\textit{e.g.} }
\def\mc#1{\mathcal{#1}}
\def\etc{\textit{etc.} }
\def\re#1{(\ref{#1})}

\newcommand{\bbb}{\bigskip}
\newcommand{\sig}{\sigma}
\newcommand{\eps}{\epsilon}
\newcommand{\ff}{\phantom{t}}
\newcommand{\ii}{\textit}
\newcommand{\bb}{\textbf}
\newcommand{\mb}{\mathbf}
\newcommand{\no}{\noindent}
\def\run{{d n_s\over d\ln k}}
\newcommand{\be}{\begin{eqnarray}}
\newcommand{\ee}{\end{eqnarray}}
\newcommand{\nn}{\nonumber}
\newcommand{\bit}{\begin{itemize}}
\newcommand{\eit}{\end{itemize}}

\def\mpc{Mpc$^{-1}$ }

\newcommand{\C}{\xi}

\newcommand{\LL}{^\ell\lambda_H}

\newcommand{\M}{m_{\mbox{\scriptsize{Pl}}}}

\def\ogw{\Omega\sub{gw}}
\def\ogwh{\omega\sub{gw}}
\begin{document}

\title{Prospects for direct detection of primordial gravitational waves.}
\author{Sirichai Chongchitnan}
\email{sc427@ast.cam.ac.uk}
\author{George Efstathiou}
\email{gpe@ast.cam.ac.uk}
\affiliation{Institute of Astronomy. Madingley Road, Cambridge, CB3 OHA. United Kingdom.}

\begin{abstract}

We study the primordial gravitational wave background produced in
models of single field inflation. Using the inflationary flow
approach, we investigate the amplitude of gravitational wave spectrum, $\omega\sub{gw}$, in the frequency range $1$ mHz - $1$ Hz pertinent to
future space-based laser interferometers. For models that satisfy the
current observational constraint on the tensor-to-scalar ratio,
$r\lesssim 0.36$, we derive a strict upper bound of $\ogwh \lesssim
1.6\times10^{-15}$ independent of the form of the inflationary
potential. Applying, in addition, the observational constraints on the
spectral index $n_s$ and its running, $\ogwh$ is expected to be
considerably lower than this bound unless the shape of the potential
is finely tuned. We contrast our numerical results with those based on
simple power-law extrapolation of the tensor power spectrum from CMB
scales.  In addition to single field inflation, we summarise a number
of other possible cosmological sources of primordial gravitational
waves and assess what might be learnt from direct detection
experiments such as LISA, Big Bang Observer and beyond.

\end{abstract}

\date{February 2006}

\maketitle

\section{Introduction}

The existence of a stochastic background of primordial gravitational
wave from inflation has yet to be verified by observation. A
significant detection would not only confirm the success of inflation,
but would also serve as a unique observational window to physics
during the very early universe. Since the first resonant bar of Joseph
Weber in the 1960s \cite{weber}, direct-detection experiments such as
LIGO have reached the stage where detection of astrophysical sources
is a realistic prospect.  Discussion of ambitious space-based
interferometers beyond LISA is well underway (see Table I for summary
and references). One of the main goals of post-LISA missions is to
detect the stochastic gravitational wave background predicted by
inflation. The most ambitious of these proposed experiments looks forward
to a precision limited only by the Heisenberg uncertainty.

In the context of inflationary models, the amplitude of the stochastic
gravitational wave background remains extremely uncertain because
neither the energy scale of inflation, nor the shape of the inflaton
potential, is known. Previous studies
\cite{turnerlett, smith} have often relied on some form of potential
to calculate the gravitational wave spectrum. While a fuller
understanding of the inflationary mechanism (if indeed inflation
occurred) awaits further development in fundamental physics, we ask
what \ii{generic} predictions, relevant to direct gravitational wave
experiments, can be made in simple models of inflation  without recourse
to specific potentials. In this paper, we address this problem and
assess the future prospects for direct detection experiments as they
confront inflation and other theoretical ideas.

After a brief overview of inflation, we calculate the amplitude of
 primordial gravitational wave spectrum predicted by inflation and
 comment on the main uncertainties involved in this calculation. We
 then generate models of inflation stochastically using the
 inflationary flow approach and study the gravitational wave
 amplitudes in these models. Direct detection experiments probe
 physical scales that are at least 15 orders of magnitude smaller than
 the scales probed by CMB experiments. The inflationary flow approach
 allows us to investigate the limitations of simple extrapolation
 between these scales using a `slow-roll' approximation. Next, we
 briefly discuss a range of other mechanisms, in addition to single
 field inflation, for generating primordial gravitational waves at
 direct detection scales.  Finally, we assess the prospects that
 future gravitational wave experiments might shed light on inflation
 and the early universe.

\begin{table*}
\begin{tabular}{|c|c|c|c|c|}
\toprule
Experiment & Time-scale &Sensitivity to $\ogwh$ & Optimum Frequency (Hz)& Reference\\
\colrule
Advanced LIGO & 2009 & $10^{-9}$ &100 & \cite{advligo}
\\
LISA &2014&$10^{-11}$& 0.005 & \cite{lisa}
\\
BBO/DECIGO&2025?&$10^{-15}-10^{-17}$& 0.1 &\cite{bbo,decigo}
\\
Ultimate DECIGO &2035?&$10^{-20}$& 0.1-1&\cite{udecigo}
\\
\botrule
\end{tabular}
\caption{
Summary of some relevant parameters of future experiments for direct
detection of gravitational waves (here
$\ogwh\equiv\Omega\sub{gw}h_0^2$).  These experiments include the
ground-based Advanced Laser Interferometer Gravitational-Wave
Observatory (LIGO), as well as space missions such as the Laser
Interferometer Space Antenna (LISA), NASA's Big Bang Observer (BBO)
and Japan's Deci-Hertz Interferometer Gravitational Wave Observatory
(DECIGO).  The ultimate DECIGO is envisaged to be a quantum limited
interferometer in space with 100-kg test masses. The quoted
time-scales and sensitivities are indicative only.}
\end{table*}

\section{Inflationary Perturbations}

We shall work in the so-called ``Hamilton-Jacobi'' formulation, in
which the Hubble parameter $H$ describes 
the inflationary dynamics. The `slow-roll' parameters $\eps$ and
$\eta$ are defined in terms of the inflaton-valued Hubble parameter
$H(\phi)$ as follows:
\be \eps \equiv {\M^2\over 4\pi}
\left({H'\over H}\right)^2\ff,\qquad\eta \equiv {\M^2\over 4\pi}\left({H''\over H}\right),\ee
where primes denote derivatives with respect to the inflaton value
$\phi$ and $\M$ is the Planck mass. Following the normalizations of
\cite{lidsey,sl}, the amplitudes $A(k)$ of primordial power spectra
$\mathcal{P}(k)$ are given, to lowest order, by
\be A_S(k) &\equiv& {2\over5} \mathcal{P}^{1/2}_S \simeq {4\over5}{H^2\over \M^2|H'|}\bigg|_{k=aH} \ff,\\
A_T(k) &\equiv& {1\over10} \mathcal{P}^{1/2}_T
\simeq{2\over5\sqrt{\pi}}{H\over \M}\bigg|_{k=aH}\ff,\label{at}\ee
where $S$ and $T$ denote scalar and tensor components respectively.
The amplitudes are evaluated when each mode, $k$, is equal in scale to
the Hubble radius, \ie when $k=aH.$ As the inflaton evolves, the rate
at which different scales leave the Hubble radius is given by
\cite{lidsey} \be {d\phi\over d \ln k} = {\M^2\over4\pi (\eps
-1)}\bigg({H'\over H}\bigg)\label{cross}.\ee

Small departures of the primordial spectra from scale invariance are
measured by the spectral indices defined as
\be n_s -1 &\equiv& {d\ln A_S^2(k)\over d\ln k} \ff,\\n_T &\equiv& {d\ln A_T^2(k)\over d\ln k} \ff.\ee In practice, however, it is common to let the spectral indices quantify variations around a pivot scale $k_0$. In this approximation,
the power spectra are parametrized by:  
\be \mathcal{P}_S(k)&=&\mathcal{P}_S(k_0)\bigg({k\over k_0}\bigg)^{n_s-1} ,\label{ps}\\ \mathcal{P}_T(k)&=&\mathcal{P}_T(k_0)\bigg({k\over k_0}\bigg)^{n_T}\ff\ff. \label{pt}
\ee Using Equation (\ref{cross}), one finds that the spectral indices can be approximated  to  $\mathcal{O}(\eps,\eta)$  by 
\be n_s -1 &\simeq& 2\eta-4\eps \ff\label{ns}, \\ n_T &\simeq& -2\eps\ff.\label{nt}\ee 
Often, it is convenient to describe a power spectrum as blue when its
index exceeds unity, or red otherwise. In this terminology, the tensor
power spectrum is said to always be tilted red. However, Pre-Big Bang
and cyclic scenarios provide exceptions, where the tensor spectrum is
strongly blue. We return to this point in Section
\ref{prospects}.

The ratio between the tensor and scalar amplitudes is clearly
\be {A_T^2\over A_S^2}\simeq \eps\ff.\ee In concordance with Ref. \cite{gpe,peiris,me}, we define the tensor-to-scalar ratio $r$ as:
\be r={\mathcal{P}_T\over\mathcal{P}_S}=16{A_T^2\over A_S^2}\simeq16\eps\ff.\label{r}\ee  
Equations (\ref{nt}) and (\ref{r}) combine to give the lowest order 
consistency relation:
\be n_T\simeq -{r\over8}\ff.\label{consist}\ee

Note that the definition of $r$ varies widely in the literature. For
instance, it is often defined as the ratio of tensor to scalar
quadrupole CMB anisotropy
${r}_2={\langle|a^T_{2m}|^2\rangle}/{\langle|a^S_{2m}|^2\rangle}$
\cite{turner, lid}. Such a definition is cosmology-dependent,
especially on the dark energy density $\Omega_\Lambda$. The conversion
is \cite{turnerwhite}:

\be {r}_2\simeq{0.84-0.025\Omega_\Lambda -0.084\Omega_\Lambda^2\over1.04-0.82\Omega_\Lambda +2\Omega_\Lambda^2}\ff r\ff.\label{convert}\ee

\section{Gravitational Wave Spectrum}\label{gwspectrum}

We now briefly derive an expression for the primordial gravitational
wave spectrum in terms of inflationary observables $r,\ff n_s$ and
$n_T$.  This Section establishes the definitions and normalizations of
various quantities used in the rest of the paper. This is important
because there are a number of derivations of the gravitational wave
energy spectrum expected from inflation in the literature, of varying
accuracy. The discussion here is based on Refs.
\cite{maggiore,buonanno,dodelson,boyle}.

We begin by considering the primordial gravitational waves produced
via tensor perturbation $h_{ij}$ of the flat
Friedmann-Robertson-Walker metric. In the synchronous gauge
($h_{\mu0}=0$) and natural units ($c=\hbar=1$), the perturbed metric
is
\be ds^2=-dt^2+a^2(t)(\delta_{ij}+h_{ij})dx^idx^j\ff,\ee 
where $a(t)$ is the scale factor in coordinate time. By further
imposing the transverse traceless conditions, the tensor perturbations
can be described by two polarization states $h_\lambda(\mb{x},t)$ with
$\lambda=+,\times$. In Fourier space, the tensor power spectrum
$P_T(k)$ observed today ($t=t_0$) is given by the variance
\be  P_T(k)\equiv{32 k^3\over \pi\M^2 }\sum_{\lambda=+,\times}\langle{h_{\lambda}^{\dagger}({k},t_0)h_{\lambda}({k},t_0)}\rangle \ff.\label{power2}\ee

Relative to the  background FRW cosmology, an effective stress-energy tensor
of gravitational waves can be defined unambiguously as \cite{isaacson}
\be T_{\mu\nu}={\M^2\over32\pi}\langle h_{ij,\mu}h^{ij}_{\ff\ff,\nu}\rangle\label{isaac}\ff.\ee 

The component $-T^0_{\ff0}=\rho\sub{gw}$ gives the energy density of gravitational wave background.  \be \rho\sub{gw}={\M^2\over32\pi}\int d(\ln k) \ff k^2 P_T(k)\ff.\label{rho}\ee

The strength of the primordial gravitational waves is characterized by the
gravitational wave energy spectrum:
\be \Omega\sub{gw}(k)={1\over\rho_c}{d\rho\sub{gw}\over d\ln k}\ff,\ee  where $\rho_c=3H_0^2/8\pi G$ is the critical density and $H_0=100h_0$ kms$^{-1}$Mpc$^{-1}$. Substituting into (\ref{rho}) gives an important result:
\be \Omega\sub{gw}(k)={1\over12H_0^2}k^2P_T(k)\ff, \label{12}\ee  which is consistent with Ref. \cite{boyle}. 
The physical density in gravitational waves is defined as 
\be \ogwh\equiv \Omega\sub{gw}h_0^2\label{little},\ee
and is independent of the value of $H_0$. Following previous work we
shall calculate constraints on the quantity $\omega\sub{gw}$.

Next, ignoring anisotropic stresses, the Einstein equations require that each state $h_\lambda(k)$ evolves via the massless Klein-Gordon equation
\be {\partial^2h_\lambda\over\partial\tau^2}+{2\over a}{\partial a\over\partial\tau}{\partial h_\lambda\over\partial\tau}+k^2h_\lambda=0\ff,\label{kg}\ee
where $\tau$ is the conformal time. Anisotropic stresses from free
streaming particles can create a non-zero source term on the
right-hand side of Equation (\ref{kg}). We return to this point
shortly.

The tensor power spectrum at the end of inflation, $\mc{P}_T(k)$,  can be 
related to the tensor power spectrum at the present day by a transfer function
$\mathcal{T}(k)$,
\be P_T(k) = \mathcal{T}^2(k)\mathcal{P}_T(k)\ff.\ee
By numerically integrating Equation (\ref{kg}),  the transfer function is found to be well approximated by
the form \cite{turner}
\no
\be \mathcal{T}(k)=
{3j_1(k\tau_0)\over{k\tau_0}}\sqrt{1.0+1.36\big({k\over
k\sub{eq}}\big)+2.50\big({k\over
k\sub{eq}}\big)^2}\ff,\label{transfer}\ee where
$k\sub{eq}=0.073\Omega_mh^2$ \mpc is the wavenumber corresponding to
the Hubble radius at the time that matter and radiation have equal energy 
densities. Using the cosmological parameters determined by combining
data from several surveys \cite{seljak}, one finds $k\sub{eq}=0.0104$
\mpc and $\tau_0=1.41\times10^{4}$ Mpc.  Combining Equations
(\ref{12}) and (\ref{transfer}) gives the gravitational wave energy
spectrum for $k\gg k\sub{eq}$:
\be\Omega\sub{gw}(k)\simeq{15\over16H_0^2 k^2\sub{eq}\tau_0^4}\mathcal{P}_T(k).\label{medium}\ee

At present, the best constraints on the normalization of the tensor
spectrum come from CMB anisotropy experiments. It is tempting
therefore to evaluate Equation (\ref{medium}) by normalizing at CMB
scales. However, the physical scales probed by CMB experiments are
about 15 orders of magnitude larger than the scales probed by direct
gravitational wave detection experiments.  In the context of this
paper, there are both positive and negative aspects associated with
this large difference in scales.  On the one hand, it is not
straightforward to extrapolate from CMB scales and infer what might be
observed by direct detection experiments, even under the restrictive
assumption of single field inflation. On the other hand, this large
difference in scales means that direct detection experiments offer the
prospect of learning something fundamentally new that cannot be probed
by CMB experiments.  The main aim of this paper is to investigate how
reliably one can extrapolate Equation (\ref{medium}) from CMB scales,
with as few constraints on the form of the inflationary potential as
possible.

Although a tensor component has not yet been observed in the CMB anisotropies,
the amplitude of the scalar component has been determined quite accurately.
At a fiducial `pivot scale', $k_0=0.002\mbox{ Mpc}^{-1}$, 
the combined results from WMAP, 2dFGRS and Lyman
$\alpha$ surveys give \cite{seljak}
\be \mathcal{P}_S(k_0=0.002\mbox{ Mpc}^{-1})\simeq2.21\times10^{-9}\ff.\label{normal}\ee
Using the above result and expressing $k$ in terms of physical
frequency $f=k/2\pi$, we finally obtain an expression for primordial
gravitational wave spectral energy in terms of $f$ and inflationary
observables $r$ and $n_T$ only:
\be\ogwh(f)\simeq4.36\times10^{-15} r\Big({f\over f_0}\Big)^{n_T}\label{large}\ff,\ee
\no where $f_0=3.10\times10^{-18}$ Hz. This relation is valid as long as $f\gg f\sub{eq}\sim10^{-17}$ Hz
and $n_T$ is independent of scale.

Further, if  $r$ and $n_T$ are accurately approximated by  first order expressions in $\eps$,
Equation (\ref{large}) becomes
\be\ogwh(f)\simeq6.98\times10^{-14}\eps\Big({f\over f_0}\Big)^{-2\eps}\ff.\label{largeeps}\ee This expression is maximized at  $\eps=[2\ln(f/f_0)]^{-1}$, with \be\ogwh(f)\big|\sub{max}\simeq{6.98\times10^{-14}\over 2e\ln(f/f_0)}\ff.\label{max}\ee 
According to this approximation, the strength of primordial
gravitational waves at direct detection scales does not increase
proportionally with $r$ because models with large $r$ have a large red
tensor tilt.  The crucial
assumption is, of course, that the power-law parametrization
$\mc{P}_T(k)\propto k^{n_T}$, with constant index $n_T$, remains
accurate over the many orders of magnitude from CMB scales to those
probed by direct detection experiments. In the next Section, we use
numerical calculations of inflationary evolution to go beyond this
approximation, finding many examples of inflationary potentials for which
Equation (\ref{large}) is violated badly.

Finally, we comment on suggestions that tensor power may be
significantly reduced  due to anisotropic stresses
from free-streaming neutrinos \cite{weinberg,bond}. For three standard
species of neutrinos, $\ogwh$ is damped by a factor of
$\lesssim(0.80)^2$ on scales which re-entered the
Hubble radius during radiation era after neutrino decoupling at a temperature
of a few MeV. These scales correspond to frequencies of about
$10^{-11}$ Hz,  well below the frequencies relevant to direct
detection of gravitational waves. Damping at direct detection
frequencies is still possible via more complicated mechanisms, for
instance, free-streaming of exotic massive particles which decouple
above the electroweak scale, or perhaps via extra dimensional physics
manifesting above the TeV scale (see Section \ref{prospects}). But because
these phenomena are still speculative and poorly understood, we have
chosen to ignore them at present. For a review of these and other
damping mechanisms, see \cite{boyle}.

\section{Numerical method}

As we have discussed above, it is interesting to analyse the stochastic
gravitational wave background  without relying on specific forms for
the inflaton potential.  Given our lack of knowledge of the
fundamental physics underlying inflation, we have tackled this problem
by investigating a large number of viable inflationary models
numerically.

Our approach  is based on the inflationary flow equations, first introduced in Ref \cite{hoffman} and further developed in \cite{gpe,me,kin,lid2,lid3,eak}. In the notation of Ref. \cite{kin}, the flow equations are:
\be {d\eps\over dN} &=& \eps(\sig+2\eps)\ff,\nn\\
{d\sig\over dN} &=& -\eps(5\sig+12\eps)+2(\ff^2\lambda_H)\ff,\\
{d\over dN}\ff^\ell\lambda_H &=& \Big[{\ell-1 \over 2}\sig +(\ell-2)\eps\Big]\ff^\ell\lambda_H+ \ff^{\ell+1}\lambda_H\ff.  \ff(\ell\geq2)\nn
\ee
Here the derivative with respect to the number of e-folds, $N$, runs
in the opposite direction to time. The flow equations represent an
infinite dimensional dynamical system whose dynamics is well
understood \cite{me}. The parameters of the system are given in terms
of inflaton-valued Hubble parameter $H(\phi)$ by:
\be \eps &\equiv& {\M^2\over 4\pi}\bigg({H'\over H}\bigg)^2,\qquad \eta \equiv {\M^2\over 4\pi}\bigg({H''\over H}\bigg)\ff,\nn\\
\ff^\ell\lambda_H &\equiv& \bigg({\M^2\over 4\pi}\bigg)^\ell{(H')^{\ell-1}\over H^\ell}{d^{\ell+1}H\over d^{\ell+1}\phi}\ff,\label{flowparam}\\
\sig &\equiv& 2\eta -4\epsilon\ff.\nn
\ee 

The hierarchy completely defines the function $H(\phi)$, which in turn determines the inflaton potential $V(\phi)$ via the Hamilton-Jacobi Equation,
\be \left(H'(\phi)\right)^2-{12\pi\over{\M^2}}H^2(\phi)=-{32\pi^2\over\M^4}V(\phi)\ff.\label{ham}\ee  
In terms of the flow parameters, the inflationary observables are
given to next to leading order by \cite{lpb}
\be r &\simeq& 16\eps[1-C(\sig+2\eps)]\ff, \label{r2}\\ n_s&\simeq&1+\sig-(5-3C)\eps^2-{1\over4}(3-5C)\sig\eps\nn\\&\ff& \ff+ \ff{1\over2}(3-C)(^2\lambda_H), \label{ns2}\\
n_T&\simeq& -2\eps - (1-C)\eps^2 +{1\over2}(1+C)\eps\sig\ff,\ee 
where $C=4(\ln2+\gamma)-5\simeq0.0814514$ (with $\gamma$ the Euler-Mascheroni constant). Variations of the spectral indices with scales are approximated to first order by the "runnings" $dn_s/d\ln k$ and $dn_T/d\ln k$. While $n_T$ may be measured directly by BBO/DECIGO (via the slope of $\omega\sub{gw}$ around 1 Hz \cite{seto}) or indirectly (via the consistency relation \cite{song,cortes}), its running, however, is likely to remain poorly constrained in the foreseeable future. Thus, we have not explicitly analysed the gravitational wave spectrum with respect to $dn_T/d\ln k$.

\begin{figure*}

\vskip 3.4 truein

\includegraphics{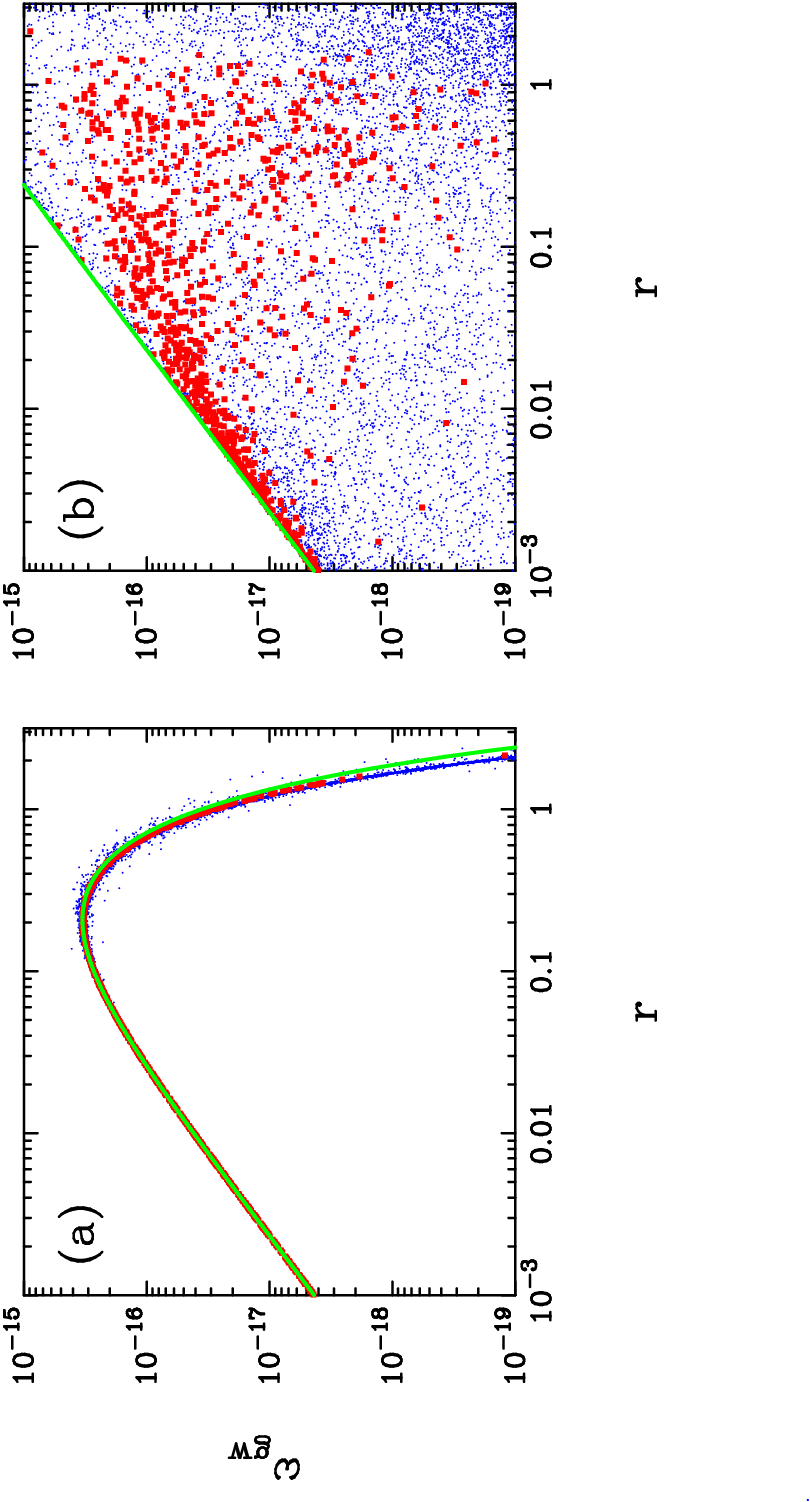}

\caption{
(Colour online) Plots of gravitational wave spectrum $\ogwh$ against
tensor-to-scalar ratio $r$ for a large number of models evolved with
the inflationary flow equations. Square (red) points indicate models
satisfying the observational constraints on $n_s$ and $dn_s/d\ln k$
given by \re{niceobs}.  In panel (a), $\ogwh$ is calculated using the
extrapolation formula \re{large}. The solid line is the first order
approximation given by Equation \re{largeeps}.
In panel (b), $\ogwh$ is calculated using the formula
\re{largest}. The solid (green)  curve in panel (b) shows
the bound given by Equation \re{upper}, with the parameter $A=7$ .}
\label{Figure1}
\end{figure*}

We ran a program (previously used in \cite{gpe,me}) that generates
models of inflation stochastically.  The program first selects the
initial configuration of a model from uniform distributions within the
following ranges:
\be \eps_0&\in&[0,0.8]\ff,\nn \\\sig_0&\in&[-0.5,0.5]\ff,\nn \\\C_0&\in&[-0.05,0.05]\ff, \label{window}\\ \LL|_0&\in&[-0.025\times5^{-\ell+3},0.025\times5^{-\ell+3}]\ff, \ff(3\leq\ell\leq10)\nn \\^{11}\lambda_H|_0&=&0\ff,\nn\ee where the hierarchy is truncated at $\ell=10$. Each model
 is evolved forward in time (backward in e-fold) until inflation ends in one of the following  ways:
\begin{enumerate}
\item By achieving $\eps=1$. When this happens, we say for convenience that the 
`slow-roll' condition has been violated. Observables on CMB scale are then 
calculated 60 e-folds before the end of inflation. This number of e-fold at which 
observables are generated is in accordance with the analyses of Refs. \cite{leach,dod}. 

\item By an abrupt termination,  perhaps from intervention of an auxiliary 
field as in hybrid inflation \cite{lr}, or, when open strings become tachyonic 
in brane inflation \cite{que,super1,anti}. Because these 
scenarios accommodate a large number of e-folds during inflation, one identifies 
them with an \ii{asymptotic} behaviour of a trajectory. In practice, those 
models inflating for more than $200$ e-folds are grouped under this category. 
The observables are then calculated along the asymptote. 
\end{enumerate}

We produced $10^6$ realizations and for each model calculated five key
observables, namely $\{r,\ff n_s, \ff n_T, \ff dn_s/d\ln k,\ff\ogwh\}$.
Working with next to leading order expressions in $\eps, \eta$, we use
the following expression for the primordial tensor power spectra with
the assumption that $\eps$ and $\eta$ are approximately constant as
each mode crosses the Hubble radius \cite{sl}
\be  \mathcal{P}_T(k)&\simeq& {16\over\pi}\left[1-({C+1\over4})\eps\right]^2 
{H^2\over\M^2}\bigg|_{k=aH},\label{nextpt}\ee where $C$ is defined as
before.  The gravitational wave spectrum depends on Equation
\re{nextpt} evaluated when the direct detection scales cross the
Hubble radius.  Since modes with frequencies in the direct detection
range of around 0.1-1 Hz exit the Hubble radius when $N\simeq20$, the
relation between the Hubble parameters at direct detection and CMB
scales is given by
\be H\sub{direct}=H\sub{CMB}\exp\left ( {-\int^{60}_{20}\eps(N)dN} \right ) \ff.\label{hint}\ee

The gravitational wave spectrum is now given in terms of the
flow-parameters at scale $k_0$ by:
\be \ogw(k)\simeq{15\over16 H_0^2k\sub{eq}^2\tau_0^4}\mc{P}_S(k_0) r\mc{I}(k)\ff, \ee 
with $r$ given by Equation \re{r2} and 
\be \mc{I}(k)=\left[{1-{C+1\over4}\eps(k) 
\over1-{C+1\over4}\eps(k_0)}\right]^2\exp\left ( -2\int_{20}^{60}\eps(N)dN
\right ) \ff.\ee  
Inserting numerical factors gives:
\be \ogwh(k)\simeq4.36\times10^{-15}r\mc{I}(k)\label{largest}\ee
For comparison between Equation \re{largest} and the extrapolation
formula \re{large}, we evaluate the gravitational wave spectrum in our
models using both expressions.  We adopted a nominal BBO/DECIGO 
frequency of $0.1$ Hz, consistent with Ref. \cite{smith}. In any case,
the results are insensitive to the choice of frequency as long as the
latter exceeds the neutrino damping scale ($\sim10^{-11}$ Hz).

\subsection{Dependence of $\ogwh$ on  $r$}

\begin{figure*}

\vskip 3.4 truein

\includegraphics{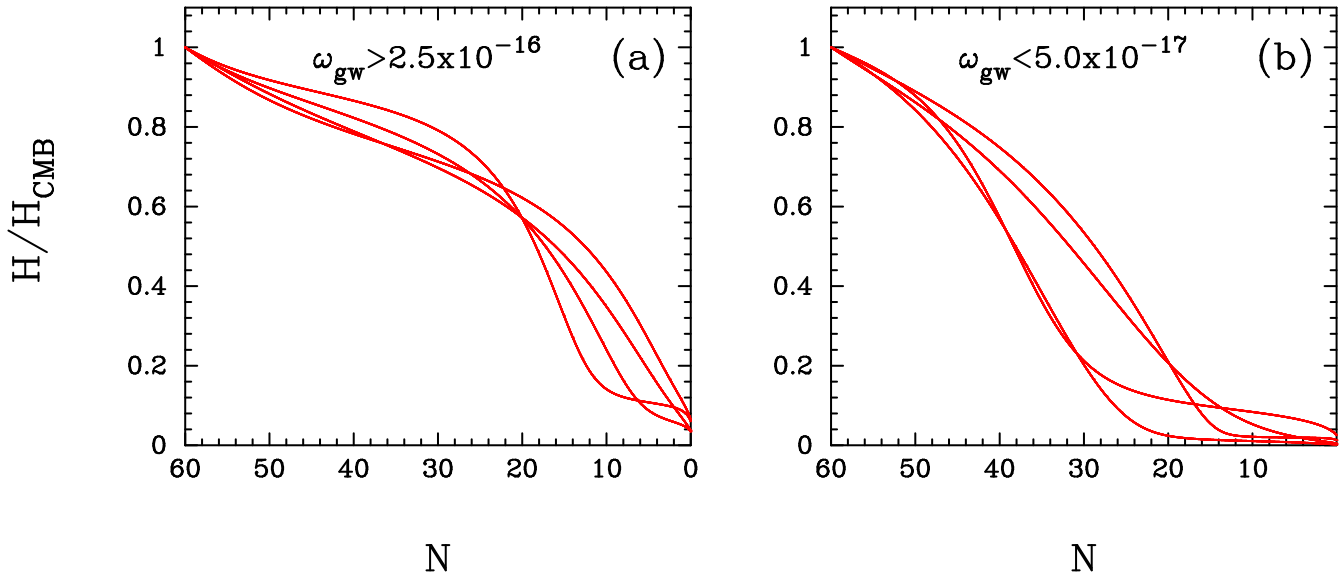}
\caption{ (Colour online) Some trajectories $H(N)$, from CMB scales ($N\simeq60$) to the end of
inflation $(N=0)$, for models evolved using the inflationary flow
equations. The models plotted in panel (a) have high gravitational
wave amplitudes at direct detection scales ($\ogwh > 2.5 \times
10^{-16}$), whilst those shown in panel (b) have low amplitudes ($\ogwh
< 5 \times 10^{-17}$).  All of these models satisfy the observational
constraints on $n_s$ and $dn_s/d\ln k$ given by Equation \re{niceobs},
and have high tensor amplitudes in the range $0.15 \le r \le 0.25$.}
\label{Figure2}
\end{figure*}

Figure \ref{Figure1} summarizes our main results. Most of the models
are of the `hybrid' type for which the tensor mode is negligible ($r
\approx 0$, $\ogwh\approx 0$) and in which the stochastic
gravitational wave background is well below the detection threshold of
any conceivable experiment.  Figure  \ref{Figure1}a shows the results
in the $r- \ogwh$ plane when the extrapolation formula
\re{large} is used  to compute $\ogwh$.  Most of the
  `non-trivial' models ({\it i.e.} models with high $\ogwh$) lie a few
percent below the first order prediction (\ref{largeeps}) shown by the
solid (green) line. All of these non-trivial models achieve $\eps=1$ at the end of inflation.
Fig. \ref{Figure1}b shows the results of using
the formula \re{largest} to compute $\ogwh$.  The distribution of
points now spans a large fraction of the $r-\ogwh$ plane.  The
inflationary flow formulation shows that the first order
extrapolation formula (\ref{large}) is too restrictive. Since the
shape of the inflationary potential is unknown, it is not possible to
extrapolate reliably from CMB scales to the much smaller scales probed
by direct detection experiments.  Figure \ref{Figure1} shows that it
is possible to find inflationary models in which, for instance, the
flow variables change rapidly within the last e-folds, thus enhancing
$\ogwh$ at direct detection scales.


The solid (green) line in Figure \ref{Figure1}b shows the expression,
\be \ogwh|\sub{max}\simeq4.36\times10^{-15} r\left[{1-{C+1\over64}Ar \over1-{C+1\over64}r}\right]^2\ff,\label{upper}\ee  where the constant $A\simeq\min\langle\eps(k_0)/\eps(k)\rangle$ depends on the distribution \re{window}. 
In our runs, we find $A\sim7$.  This expression provides an accurate
upper bound to $\ogwh$. Equation
\re{upper} simply expresses the constraint that the Hubble parameter
is constant between CMB and direct detection scales, modulated by the
term in square brackets which expresses the details of how inflation
ends.  However, for any value $r \lesssim 1$, the term in square
brackets is close to unity and so is insensitive to the parameter $A$
and hence to the distribution \re{window}.

The (red) square points in Figure \ref{Figure1} show the subset of
models that satisfy the $2\sigma$ observational
constraints \cite{seljak,matteo} on $n_s$ and $dn_s/d\ln k$,
\begin{equation}
0.92\lesssim n_s
\lesssim 1.06,  \quad -1.04\lesssim \run\lesssim 0.03.  \label{niceobs} 
\end{equation}
These models roughly follow the 
locus of the first order extrapolation shown in Figure \ref{Figure1}a, but
with a large scatter. As a conservative bound we apply Equation
\re{upper} with the observational constraint $r<0.36$ \cite{seljak}, to give
\be\ogwh\lesssim1.6\times 10^{-15}.\label{wmax}\ee 

\begin{figure*}

\vskip 3.4 truein

\includegraphics{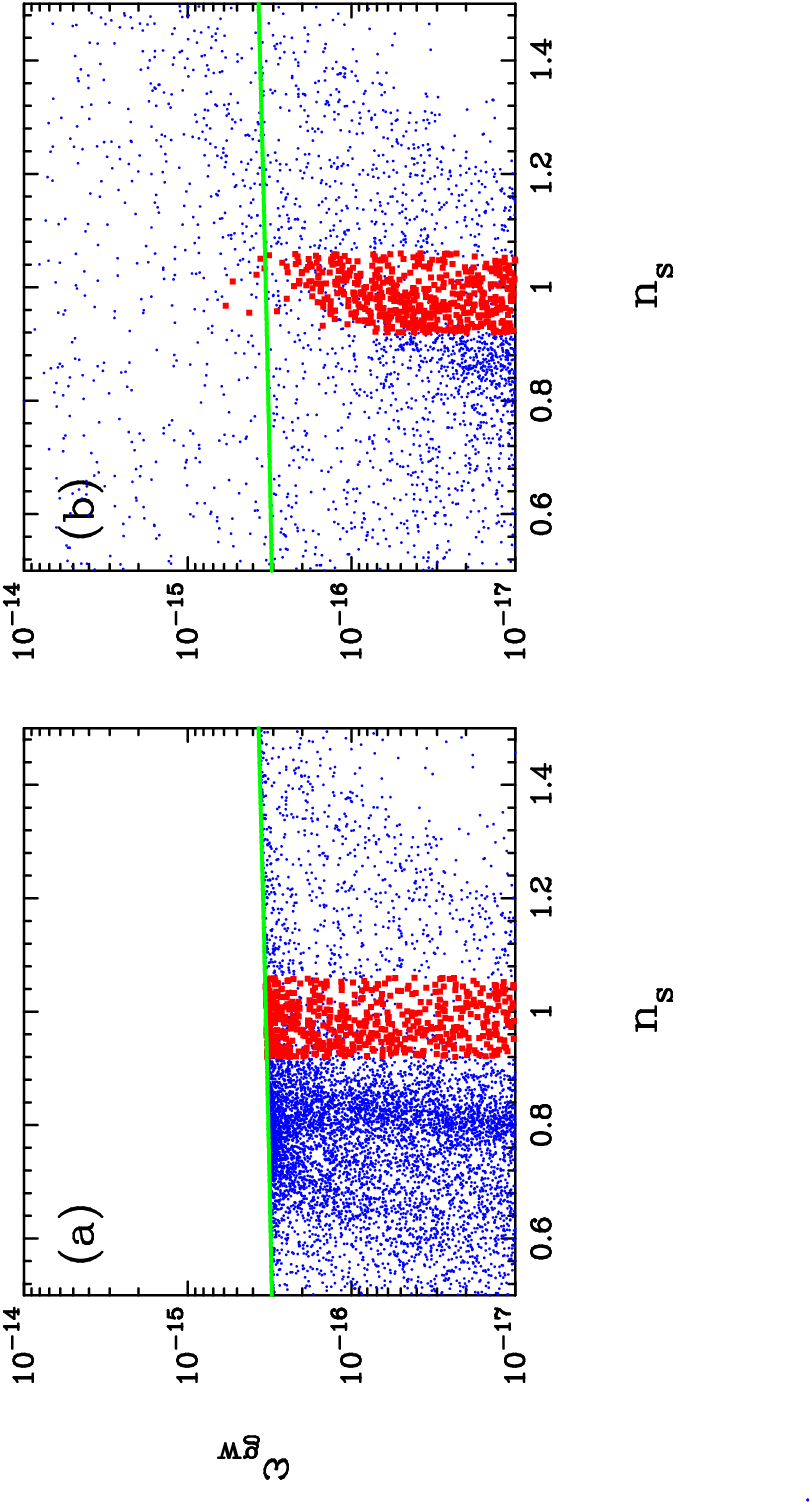}
\caption{ (Colour online) The gravitational wave spectrum $\ogwh$ 
plotted against scalar spectral index $n_s$ for a large number of
models evolved using the inflationary flow equations. Square (red)
points indicate models satisfying the observational constraints on
$n_s$ and $dn_s/d\ln k$ (Equation \re{niceobs}) and satisfying
$r<0.36$.  In panel (a), $\ogwh$ is calculated using the extrapolation
formula \re{large}.  The (green) solid curve shows the bound given by
Eq. \re{env}.  In panel (b), $\ogwh$ is calculated using formula
\re{largest} and the flow equation integration. }
\label{Figure3}
\end{figure*}

As this paper was nearing completion, a paper by \cite{smith2} appeared describing
a similar analysis. Our results are broadly compatible, but
there appear to be some discrepancies. Comparing our Figure
\ref{Figure1}b with their Figure 2, we see that the swathe of points
satisfying \re{niceobs} matches roughly the shape of the contoured
region in their Figure. However, we find models with low values of
$\ogwh \lesssim 10^{-18}$ at all values of $r$ whereas they do
not. Furthermore, at high values of $r \gtrsim 0.1 $, they appear to
find models that lie above the bound given by \re{upper}. Their
results do not seem physically plausible to us \footnote{This is because Ref. \cite{smith2} extrapolates from CMB scales to direct detection scales (using $n_T$ and its running) in order to test the consistency relation. We thank Hiranya Peiris for clarification. }.

Examples of some trajectories $H(N)$, from CMB scales to the end of
inflation, are shown in Figure \ref{Figure2}. All of these models
satisfy the observational constraints on $n_s$ and $dn_s/d\ln k$ of
Equation
\re{niceobs} and, in addition, we have imposed the constraint 
$0.15 \le r \le 0.25$, {\it i.e.} the models have high tensor
amplitudes. The models plotted in Figure \ref{Figure2}a have high
gravitational wave amplitudes at direct detection scales ($\ogwh > 2.5
\times 10^{-16}$). In these cases, the Hubble parameter stays almost
constant from $N=60$ to $N=20$ but declines rapidly thereafter.
In contrast, the models shown in Figure
\ref{Figure2}b have low amplitudes $\ogwh < 5 \times 10^{-17}$. 
In these cases, $H(N)$ declines more rapidly between $N=60$ and
$N=20$. These sample trajectories show that models with sharp features
in $H(N)$ (and hence also in $V(\phi)$) within the last $20$ e-folds
of inflation will be the first to be ruled out by BBO/DECIGO-type
detectors.

\subsection{Dependence of $\ogwh$ on scalar tilt $n_s$}

Figure \ref{Figure3} shows the models plotted in the $n_s-\ogwh$
plane.  The extrapolation method (Fig. \ref{Figure3}a) places most 
of the `non-trivial' models within a vertical band centered around
$n_s\sim0.8$. The band is sharply capped by the solid (green) curve
given by differentiating Eq.\re{large}:

\be \ogwh|\sub{max}= 4.24\times10^{-17}\left[{17.235-1.303n_s\over2.565-0.541n_s}\right],\label{env}\ee

Fig.\ref{Figure3}b shows the distribution when the flow formulation
\re{largest} is used to calculate $\ogwh$. The region beyond the
envelope \re{env} is now populated by many models, some of which
produce $\ogwh$ in excess of $10^{-14}$.  However, all of the models
with such high values of $\ogwh$ are inconsistent with the
observational constraints on $n_s$ and $d n_s/ d\ln k$.  The (red) squares in
Figure \ref{Figure3} indicate models that satisfy the $2\sigma$
observational constraints of Equation \re{niceobs}, and, in addition,
have $r<0.36$.  The vast majority of these models lie below the line
defined by Equation \re{env}. However, it is possible, though rare, for models satisfying the
observational constraints \re{niceobs} to exceed $\ogwh > 3 \times
10^{-16}$, as given by Equation \re{env}. Evidently, one can see from Figure \ref{Figure1}b that no model
satisfying the observational constraints can exceed our conservative
bound \re{wmax}.

\section{Prospects for direct detection}\label{prospects}

The results of the preceding Section show that simple single-field
inflation models must satisfy the conservative constraint of Equation
\re{wmax} at direct detection scales. Furthermore, unless the
inflationary parameters are specially tuned, most single-field
inflation models will produce $\ogwh \lesssim3 \times 10^{-16}$. Thus,
at the BBO/DECIGO sensitivities of $\sim 10^{-15}-10^{17}$ (see Table 1), a direct
detection of a stochastic background of gravitational waves would be
expected only if the inflationary potential contains a feature at $N
\sim 20$,  as shown in the trajectories plotted in Figure \ref{Figure3}.
This is true even if the tensor-to-scalar ratio is high at CMB scales. 
This is the main conclusion of this paper.

Although this may seem a somewhat pessimistic conclusion for direct
detection experiments, it is worth mentioning a range of other
cosmological sources (summarized in Table II) that could produce a
stochastic background of gravitational waves at direct detection scales.  

\medskip 

\noindent
\ii{(i) Pathological potential:} A
sudden decrease in energy scale of the inflationary universe could be
attributed to a first order phase transition brought about by the
spontaneous symmetry breaking of a field coupled to the inflaton. As a
result, the potential $V(\phi)$ also acquires a sharp feature in the
form of steps \cite{double,doubles,park,adams}, kinks
\cite{starobinsky,lesga,lesgb} or combination of these at various
scales \cite{ars}. In particular, the primordial gravitational wave 
amplitude in the so-called `broken-scale invariance' models has been
considered in Refs. \cite{smith,polarski}, which found roughly an
order of magnitude increase above that  given by Equation
\re{largeeps}. Clearly, a first order phase transition has a
negligible enhancement effect on modes at direct detection scales
unless the transition occurs at late stages (within the last $\sim$20
e-folds) of inflation.  On the other hand, if scale invariance is broken
at around the CMB/LSS scales, as suggested by
\cite{Mukherjee,einasto,gaztanaga}, then the gravitational wave
amplitude may be enhanced at scales probed by the future CMB
polarization experiments.

\medskip 

\noindent
\ii{(ii) Bubble nucleation:} A phase transition may also be accompanied
by a rapid nucleation of vacuum bubbles
\cite{highest,extended1,extended2}, which upon collision during
inflation produce a large gravitational wave background with $\ogwh$
of order $\sim10^{-7}$ around the direct detection
frequencies. However, bubble collision at a much lower energy, \eg the
electroweak scale, produces virtually negligible gravitational waves
with $\ogwh$ of order $10^{-23}$ \cite{turbo}. In supersymmetric
extentions of the standard model, this value may be larger by several 
orders of magnitude \cite{mssm} and perhaps as large as $\sim
10^{-11}$ for some parameter choices in next-to-minimal models.

\medskip

\noindent
\ii{(iii) Turbulence:} A large injection of energy into the cosmological
plasma following bubble collision could also set up a Kolmogorov
spectrum of turbulence. Calculations in Refs. \cite{turbo,turbo2}
estimate the gravitational wave background from turbulence to be
comparable to that from bubble nucleation. If the turbulence is
sourced also by a helical field (\eg primordial magnetic fields), a
secondary contribution of $\ogwh\sim10^{-11}$ is predicted at direct
detection scales \cite{ratra}. Relation between $\ogwh$ and the  
strength of primordial magnetic fields is further discussed in \cite{durrer1,durrer2}

\medskip
\noindent
\ii{(iv) Cosmic strings:} 
A stochastic network of strings \cite{sv, kibble} produces a
gravitational wave spectrum with a long plateau extending from
$f\sim10^{-10}$ Hz across direct detection scales \cite{u1,u2}.
Although CMB observations show that strings cannot be solely
responsible for structure formation \cite{wyman}, they can arise in
certain models of hybrid and brane inflation as a sub-dominant
contribution to the fluctuations \cite{super1,super2,Firouzjahi}.
Recently Refs. \cite{cusp1,cusp2} have calculated the gravitational
wave spectrum from bursts associated with cusps and kinks in loops of
cosmic (super)strings as a function of the theoretically uncertain
intercommutation probability.  They conclude that the gravitational wave bursts from strings with tensions as low as 
$G\mu\sim10^{-14}$ could result in $\ogwh$ as large as $\sim10^{-11}$. This is potentially detectable by LISA and may even be observable by LIGO if $G\mu\gtrsim10^{-10}$ and the intercommutation probablity small.

\medskip

While inflation may be accompanied by all of the phenomena mentioned
above, some alternatives to slow-roll inflation have altogether
different predictions regarding the production of primordial
gravitational waves at direct detection scales.

\medskip

\noindent
\ii{(v) Pre-Big Bang and cyclic models:} In Pre-Big Bang scenarios
\cite{sdual,vene}, a dilaton-driven phase with $\dot{H}>0$ gives rise
to a gravitational wave amplitude which increases with frequency
($\sim f^3$) for all modes exiting the Hubble radius during the
Pre-Big Bang era. The primordial tensor spectrum in this case is
strongly blue with $n_T=3$. The gravitational wave spectrum could peak
at direct detection scales with amplitude $\ogwh$ as high as
$10^{-6}$, within reach of advanced terrestrial detectors
\cite{pbb1,pbb2,pbb3}. When combined with CMB polarization
experiments, a strongly blue tensor spectrum can be easily ruled
out. Nevertheless, the prediction of such a large gravitational wave
amplitude at direct detection scales is sensitive to physics during
the `bounce' around $t=0$, which remains poorly understood
\cite{review}. In contrast, the cyclic model
\cite{cyclic,cyclicbig} predicts a blue tensor spectrum ($n_T=2$) but
with negligible gravitational wave amplitude at direct detection
scales \cite{cygw}.  

\medskip

\noindent
\ii{(vi) Braneworlds:} Inflation has been implemented in
5-dimensional phenomenological braneworld models
\cite{rs1,rs2,trap}. Gravitational waves at direct detection scales cross the Hubble radius at high energies ($H\ell\gg1$, where $\ell$ is the bulk curvature), hence $\ogwh$ is directly affected by extra-dimensional physics. An enhancement effect in $\ogwh$ arises through the modification of the Friedmann equation, whereas a damping effect occurs via 
the mixing of massive Kaluza-Klein modes with the massless graviton
\cite{hira1,mix,ichiki}. At direct detection scales, it is conceivable that
these two effects cancel \cite{hira2, cancel}.

\begin{table*}
\begin{tabular}{|p{3.8cm}|l|c|c|c|}
\toprule
\qquad\ff\ff\ff\ff Phenomena & Key parameters & $\ogwh$(1 mHz - 1 Hz) &  References\\
\colrule
1. Slow-roll inflation & Inflationary energy scale.   & $\lesssim10^{-15}$ & Eq. \re{wmax}
\\

&&&\\

2. Pathological potential &  \begin{minipage}[t]{5.4cm} \flushleft `Breaking' scale(s).\\ Sharp changes in $V, V' , V''$ \etc \\ \end{minipage}&$\lesssim10^{-15}$\ff& \cite{smith,polarski}
\\
&&&\\

3. Bubble nucleation &   \begin{minipage}[t]{5.4cm}  \flushleft Bubble velocity.  \\Energy scale of transition. \\Time scale of transition. \\Efficiency of energy conversion. \\ + SUSY parameters \end{minipage} & \begin{minipage}[t]{2.4cm} $\lesssim10^{-7}$\\$10^{-11}-10^{-16}$\\$10^{-23}$ \end{minipage}&\begin{minipage}[t]{2cm} \cite{highest,extended1,extended2} \\ \cite{mssm} \\ \cite{turbo}\end{minipage}
\\
&&&\\
4. Turbulence &\begin{minipage}[t]{5.4cm}  \flushleft Characteristic turbulent scale.\\ Damping scale.\\ Energy scale of turbulence.\\ Time scale of turbulence.  \\ Efficiency of energy conversion.\\ + Detail of helical/magnetic fields \\ \end{minipage}& \begin{minipage}[t]{2.4cm} $\lesssim10^{-7}$\\$10^{-12}$\end{minipage}&\begin{minipage}[t]{2cm} \cite{turbo,turbo2}\\ \cite{ratra} \end{minipage}
\\
&&&\\

5. Cosmic strings & \begin{minipage}[t]{5.4cm} \flushleft String tension. \\ Average loop size. \\ Intercommutation probability. \\ Burst rate. \\ + Detail of loop distribution\\\end{minipage}&\begin{minipage}[t]{2.4cm} $10^{-9}-10^{-11}$\end{minipage}&\begin{minipage}[t]{2cm} \cite{cusp2} \end{minipage}
\\

&&&\\
\colrule
6. Pre-Big Bang / Cyclic \ff\ff\ff models& \begin{minipage}[t]{5.4cm} \flushleft Detail of  stringy epoch?  \\\end{minipage}&\begin{minipage}[t]{2.4cm}$\lesssim10^{-6}$\\$10^{-35}$\end{minipage}& \begin{minipage}[t]{2cm} \cite{pbb1,pbb2,pbb3}\\ \cite{cygw}\end{minipage}

\\
&&&\\
7. Braneworlds  &\begin{minipage}[t]{5.4cm} \flushleft Bulk curvature scale.  \\  Radion potential.\\  + Geometrical setup \end{minipage}& $\lesssim10^{-16}$&\begin{minipage}[t]{2cm} \cite{ichiki,cancel}\end{minipage}
\\
\botrule
\end{tabular}
\caption{Summary of some possible cosmological sources of primordial gravitational wave background in the frequency range of future direct detection experiments ($f\simeq$ 1 mHz - 1 Hz). Inflation may be accompanied by some (or all)
of phenomena 2-5, while phenomena 6 and 7 are alternatives to the inflationary scenario.}
\end{table*}

\medskip 
Finally it is worth noting that we have ignored astrophysical sources,
most notably from inspiralling binary systems of white dwarfs, neutron
stars or black holes which could produce a significant background at
frequencies of 1 mHz to $1$ Hz. These sources must be subtracted to
high accuracy \cite{astro, cooray, harms} to achieve sensitivities of $\ogwh
\ll 10^{-15}$ necessary to test inflation, and may ultimately limit
direct detection experiments. The sensitivities of the post-LISA
experiments quoted in Table I depend on the usable frequency range
and are significantly lower if frequencies $\lesssim 0.2$ Hz are
contaminated by a high background from unresolved white dwarfs.

\section{Conclusions}

The generation of tensor modes is a key prediction of inflationary
models and has yet to be confirmed by experiment.  A large
experimental effort is underway to detect a tensor mode signature in
the polarization of the CMB. On a longer time-scale, a number of
direct detection experiments have been proposed to detect a stochastic
background of gravitational waves at frequencies in the range 1 mHz -
1 Hz. However, since the spatial scales probed by direct experiments
are some $15$ orders of magnitude smaller than the scales probed by
the CMB, extrapolating between these scales is highly model dependent
\cite{turnerlett,smith}.

In this paper, we have used the inflationary flow equations to assess
the accuracy of extrapolating between CMB and direct detection scales
for single-field inflationary models. Our main results are shown in
Figures 1 and 3. For models that satisfy the observational constraints
on $n_s$ and $dn_s/d\ln k$, we find a conservative upper bound of $\ogwh
\lesssim 1.6 \times 10^{-15}$. However, as shown in Figure
\ref{Figure3}b most of our models have much lower values
of $\ogwh$, and only a small minority have $\ogwh \gtrsim 3 \times
10^{-16}$. A direct detection experiment with a sensitivity of $\ogwh
\sim 10^{-16}$ is therefore limited to testing a range of
single field inflationary models in which the Hubble parameter is
roughly constant between CMB scales ($N \approx 60$) and direct
detection scales ($N \approx 20$), followed by an abrupt decline
thereafter. Examples of such trajectories are shown in Figure
\ref{Figure2}a.

We have also identified a number of cosmological sources of stochastic
gravitational wave background accessible to direct detection
experiments (Table II). In some cases, the predicted amplitudes are
far in excess of those generated during inflation.  A high value of
$\ogwh$ from, say, cosmic strings produced at the end of brane
inflation might easily overwhelm the contribution from tensor modes
generated during inflation. In more general scenarios, therefore, it
may be difficult for direct detection experiments to constrain the inflationary phase
 even if experiments can achieve `Ultimate DECIGO' sensitivities of $\ogwh
\sim 10^{-20}$.

\bbb

\no \bb{Acknowledgments:} We thank Antony Lewis for useful discussions. SC acknowledges the support of a Dorothy Hodgkin scholarship from PPARC.  This work has been supported by PPARC.

\end{document}